\documentclass[%
 reprint,
superscriptaddress,
 amsmath,amssymb,
 aps,
]{revtex4-2}
\usepackage{graphicx}
\usepackage{amsfonts}
\usepackage{amsthm}
\usepackage{multirow}
\usepackage{color}
\def\beq{\begin{equation}}
\def\eeq{\end{equation}}
\def\bea{\begin{eqnarray}}
\def\eea{\end{eqnarray}}
\newcommand{\beqs}{\begin{subequations}}
\newcommand{\eeqs}{\end{subequations}}

\newcommand{\cref}[1]{Ref.~\cite{#1}}

\newcommand{\hh}{{\ensuremath{I{\kern-2.6pt h}}}}
\newcommand{\bhh}{{\ensuremath{\bar{I{\kern-2.6pt h}}}}}

\usepackage[colorlinks=true,allcolors=blue]{hyperref}

\usepackage{graphicx}
\usepackage{dcolumn}
\usepackage{bm}

\begin{document}

\title{Superheavy quasi-stable strings and walls bounded by strings in the light of NANOGrav 15 year data}
	\author{George Lazarides}
	\affiliation{School of Electrical and
Computer Engineering, Faculty of Engineering, Aristotle University of Thessaloniki, Thessaloniki 54124, Greece}
	\author{Rinku Maji}
	\affiliation{Laboratory for Symmetry and Structure of the Universe, Department of Physics,
		  Jeonbuk National University, Jeonju 54896, Republic of Korea}
	\author{Qaisar Shafi}
	\affiliation{Bartol Research Institute, Department of Physics and 
		Astronomy,
		 University of Delaware, Newark, DE 19716, USA}

\begin{abstract}
Composite topological structures such as superheavy ``quasi-stable strings" (QSS) and ``walls bounded by strings" (WBS)
arise in realistic extensions of the Standard Model of high energy physics. We show that the gravitational radiation emitted in the early universe by these two unstable structures with a dimensionless string tension $G\mu\approx 10^{-6}$ is consistent with the NANOGrav evidence of low frequency gravitational background as well as the recent LIGO-VIRGO constraints, provided the superheavy strings and monopoles experience a certain amount of inflation. For the case of walls bounded by strings, the domain walls arise from the spontaneous breaking of a remnant discrete gauge symmetry around the electroweak scale. The quasi-stable strings, on the other hand, arise from a two step breaking of a local gauge symmetry. The monopoles appear from the first breaking and get connected to strings that arise from the second breaking. Both composite structures decay by emitting gravitational waves over a wide frequency range. The Bayes factors for QSS and WBS relative to the inspiraling supermassive black hole binaries are estimated to be about 60 and 30 respectively, which are comparable with that of metastable strings and cosmic superstrings.

\end{abstract}

\maketitle
\section{Introduction}
The recently reported evidence of a stochastic gravitational background radiation by the NANOGrav \cite{NANOGrav:2023gor,NANOGrav:2023hvm} and other pulsar timing array collaborations \cite{Antoniadis:2023ott,Reardon:2023gzh,Xu:2023wog} provides strong impetus for testing the predictions of particle physics models containing strings, walls as well as composite topological structures. For instance, topologically stable cosmic strings with dimensionless string tension $G \mu \gtrsim 10^{-10}$ ($G$ is Newton's gravitational constant and $\mu$ is the string tension) appear to be excluded by the latest NANOGrav data. A number of articles discussing the NANOGrav data have recently appeared \cite{NANOGrav:2023hfp,Franciolini:2023wjm,Shen:2023pan,Zu:2023olm,Lambiase:2023pxd,Han:2023olf,Guo:2023hyp,Wang:2023len,Ellis:2023tsl,Vagnozzi:2023lwo,Fujikura:2023lkn,Kitajima:2023cek,Li:2023yaj,Franciolini:2023pbf,Megias:2023kiy,Ellis:2023dgf,Bai:2023cqj,Yang:2023aak,Ghoshal:2023fhh,Deng:2023btv}.

In this paper, we would like to point out that well-known composite topological structures, known as ``Quasi-Stable Strings (QSS)" \cite{Lazarides:2022jgr} and ``Walls Bounded by Strings (WBS)" \cite{Kibble:1982dd}, which emit \cite{Martin:1996ea} background gravitational radiation, can be consistent with the NANOGrav evidence. In both cases, the strings are superheavy with a dimensionless string tension $G\mu \sim 10^{-6}$.
In the case of QSS, the structure consists of superheavy monopoles at the ends of strings, and it disappears after the gravitational wave emission by the strings. Note that the monopoles experience a period of inflation before getting attached to the strings and reentering the horizon. In contrast to metastable strings \cite{Preskill:1992ck}, the monopole-antimonopole pair creation on the string is not effective in the QSS scenario.
In WBS, the domain walls arise from the spontaneous breaking of a discrete gauge symmetry at the electroweak scale.

Since the strings radiate gravitational waves over a wide frequency range, if they are superheavy they may not be compatible with the third run advanced LIGO-VIRGO \cite{LIGOScientific:2021nrg} results.
Indeed, the superheavy strings that form boundaries of the electroweak scale domain walls or connect the superheavy monopoles should experience a certain amount of inflation in order to avoid this conflict. A rough estimate shows that the strings should experience partial inflation and reenter the horizon at a cosmic time of order $10^{-10}$ sec or later.

We do not intend here to provide details on how these topological structures may come about, but it is perhaps worth mentioning specific realistic particle physics models where they could be implemented.
For an example of how the QSS structures arise, consider the following breaking of flipped $SU(5)$ \cite{Lazarides:2023iim}:
\begin{align}
& SU(5) \times U(1)_X  \nonumber \\ &\xrightarrow[]{\left<24(0)\right>}  SU(3)_C \times SU(2)_L \times U(1)_Z \times U(1)_X \nonumber \\
& \xrightarrow[]{\left<10(-1)\right>} SU(3)_C\times SU(2)_L\times U(1)_Y.
\end{align}
We have listed the $SU(5)\times U(1)_X$ representations responsible for the breakings by their appropriate vacuum expectation values.
The first breaking yields superheavy monopoles carrying color, electroweak and $U(1)_Z$ magnetic charges. The subsequent breaking of $U(1)_Z \times U(1)_X$ to $U(1)_Y$ at a somewhat lower scale reveals that the monopoles are topologically unstable and get connected to strings, which we call QSS. These monopoles do not carry unconfined fluxes after the electroweak breaking.

For another example, consider the following breaking of $SO(10)$ \cite{Lazarides:2023iim}:
\begin{align}
SO(10) &\xrightarrow[]{\left<45\right>} SU(3)_C \times SU(2)_L \times U(1)_Y \times
U(1)_\chi \nonumber \\
&\xrightarrow[]{\left<16\right>} SU(3)_C \times SU(2)_L \times U(1)_Y .        
\end{align}
The first breaking yields the standard $SU(5)$ monopoles (which  
presumably will be adequately diluted by inflation) as well as superheavy  
monopoles carrying  $U(1)_\chi$ and $U(1)_Y$ magnetic
charges. The subsequent breaking of $U(1)_\chi$ at a somewhat lower scale  
reveals that the latter monopoles are topologically unstable and get  
connected to strings (QSS). Note that after the  
electroweak symmetry breaking these monopoles do not carry any Coulomb  
magnetic flux.

 To see how the WBS system appears, consider the following breaking of the flipped $SU(5)$ model \cite{Lazarides:2023iim}:
\begin{align}\label{eq:f-su5}
&SU(5)\times U(1)_X \nonumber \\
&\xrightarrow[]{\left<50(2)\right>}SU(3)_C\times SU(2)_L\times U(1)_Y \times{Z}_2 \nonumber\\
&\xrightarrow[]{\left<\overline{10}(1)\right>} SU(3)_C\times SU(2)_L\times U(1)_Y.
\end{align}
The first breaking produces topologically stable superheavy strings but no monopoles. A subsequent breaking of the gauge $Z_2$ symmetry yields the desired composite structures, namely ``walls bounded by strings."
\section{Gravitational wave background and NANOGrav 15 year data}
\label{sec:GWs}
The strings start forming loops through inter-commuting after the horizon reentry of the string network at a time $t_F$. This network behaves like a network of stable strings before the horizon reentry of the monopoles at time $t_M$ for the QSS, and before the time $R_c=\mu/\sigma$ ($\sigma$ is the wall tension) for the string-wall network. After $t_{\rm fin} = t_M$, there will be contribution from the loops formed earlier and the monopole-antimonopole pairs connected by string segments ($MS\bar{M}$) present within the particle horizon. 
On the other hand, WBS structures larger than $R_c$ will collapse and, therefore, only loops of size less than $R_c$ formed before the time $t_{\rm fin}=R_c$ contribute to the gravitational waves after $R_c$. In both cases, the network disappears after a time scale $\sim$ $t_{\rm fin}/\Gamma G\mu$ ($\Gamma\sim 10^2$) \cite{Vilenkin:1982ks,Vachaspati:1984gt,Vilenkin:2000jqa} and the bound from the cosmic microwave background anisotropy \cite{Planck:2013mgr,Charnock:2016nzm} is alleviated.

\begin{figure}[htbp]
\begin{center}
\includegraphics[width=0.95\linewidth]{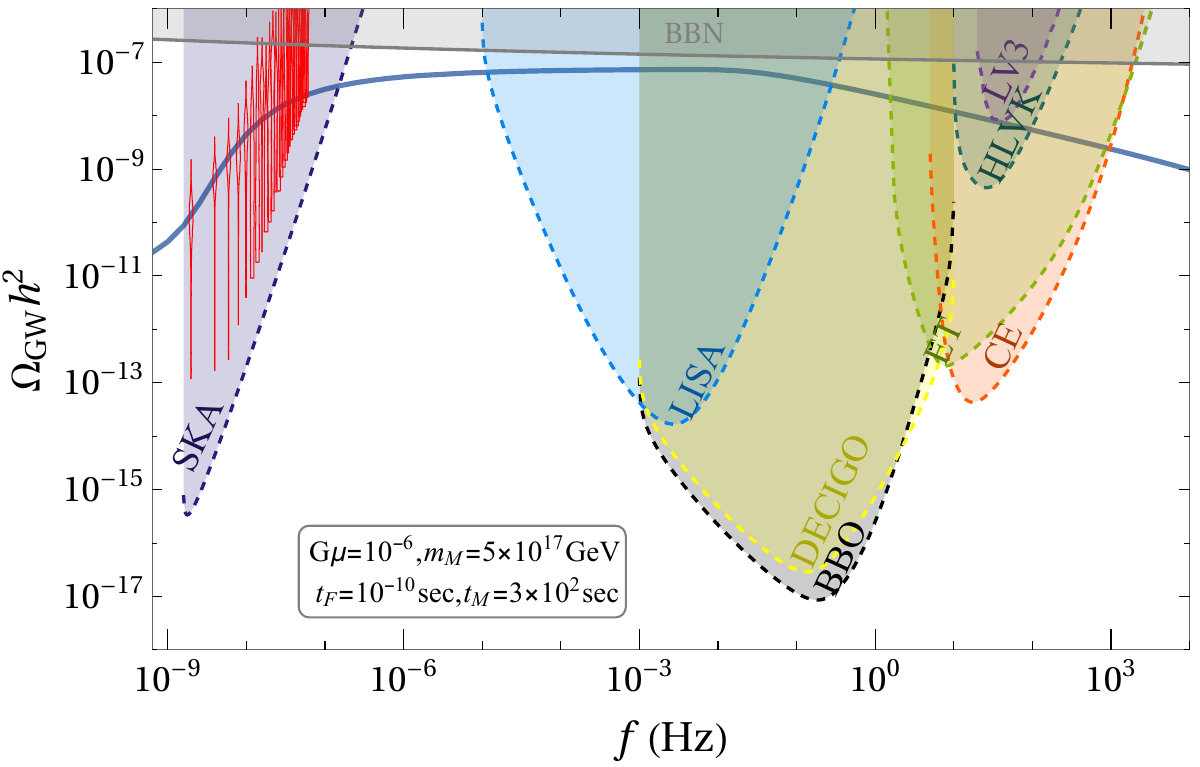}
\end{center}
\caption{Gravitational wave background from quasi-stable strings with $G\mu = 10^{-6}$ and the monopole horizon reentry time $t_M=3\times 10^2$ sec. The evidence for gravitational waves in NANOGrav is compatible with this scenario.  The red violin plots show the posterior of HD correlated free spectra of PTA data. The strings experience some $e$-foldings and reenter the horizon at $t_F\sim 10^{-10}$ sec to satisfy the advanced LIGO-VIRGO third run (LV-3) bound \cite{LIGOScientific:2021nrg}. The gray region depicts the bound from Big Bang Nucleosynthesis (BBN) \cite{Mangano:2011ar}. We also show the power-law integrated sensitivity curves \cite{Thrane:2013oya, Schmitz:2020syl} for planned experiments, namely, HLVK \cite{KAGRA:2013rdx}, CE \cite{Regimbau:2016ike}, ET \cite{Mentasti:2020yyd}, DECIGO \cite{Sato_2017}, BBO \cite{Crowder:2005nr, Corbin:2005ny}, LISA \cite{Bartolo:2016ami, amaroseoane2017laser}, and SKA \cite{5136190, Janssen:2014dka}.}\label{fig:GWs-QSS}
\end{figure}

\begin{figure}[htbp]
\begin{center}
\includegraphics[width=0.95\linewidth]{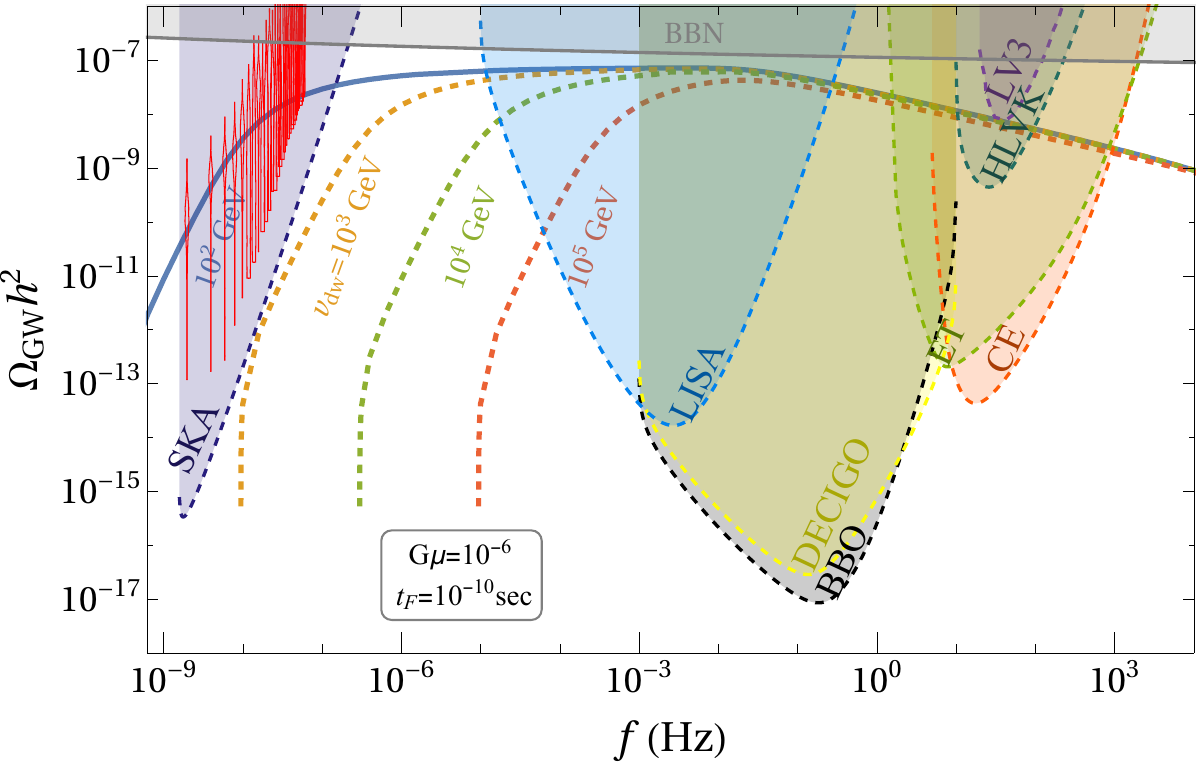}
\end{center}
\caption{Gravitational wave background from domain walls bounded by cosmic strings with $G\mu = 10^{-6}$ and different choices for the vacuum expectation value $v_{\rm dw}$ associated with the domain walls. The red violin plots show the posterior of  HD correlated free spectra of PTA data. The NANOGrav evidence is compatible with this scenario for $v_{\rm dw}\sim 10^2$ GeV. The strings experience some $e$-foldings of inflation, and reenter the horizon at $t_F\sim 10^{-10}$ sec to satisfy the bound from the third run advanced LIGO-VIRGO (LV-3) data \cite{LIGOScientific:2021nrg}.}\label{fig:GWs-DWS}
\end{figure}

 We follow the prescription in Ref.~\cite{Lazarides:2022jgr} to compute the gravitational wave spectra from the QSS, and take into account that the monopoles do not carry Coulomb magnetic flux. The gravitational wave background from the WBS network is computed following Refs.~\cite{Dunsky:2021tih,Maji:2023fba}. The gravitational wave background from the string loops is the sum of the contributions from all the normal modes:
\begin{align}
\Omega_{\rm GW}(f) = \sum_{k=1}^{\infty}\Omega_{\rm GW}^{(k)}(f) .
\end{align}
The contribution from each mode in the case of WBS is given by \cite{Dunsky:2021tih,Maji:2023fba}
\begin{align}
   \label{eq:GWs2}
    \Omega^{(k)}_{\rm GW}(f) 
    = &\frac{1}{\rho_c} \int_{t_F}^{t_0} d\tilde{t} \left(\frac{a(\tilde{t})}{a(t_0)}\right)^5\frac{\mathcal{F} C_{\rm eff}(t_i)}{\alpha t_i^4} \left(\frac{a(t_i)}{a(\tilde{t})}\right)^3 \nonumber \\
    & \frac{\left(1 + \frac{2 k}{2\pi R_c f}\frac{a(\tilde{t})}{a(t_0)} \right)}{\Gamma G \mu + \alpha(1 + \frac{\alpha t_i}{2\pi R_c})}\frac{\Gamma k^{-4/3}}{\zeta(4/3)} G\mu^2 \frac{2 k}{f}\Theta(t_{\rm fin} - t_i) ,
\end{align}
and a loop formed at time $t_i$ has length $l$ at a subsequent time $t$ given by
\begin{align}
    \label{eq:lengthLoss}
    G\mu(t - t_i) = \int_l^{\alpha t_i} \frac{dl'}{\Gamma}\left(1 + \frac{l'}{2 \pi R_c}\right).
\end{align}
Here, $\rho_c$ is the critical energy density at the present time $t_0$, $a(t)$ the scale factor of the universe, $\mathcal{F}\simeq 0.1$, $\Gamma \simeq 50$, $\alpha\simeq 0.1$ in the pure string limit \cite{Vachaspati:1984gt,Vilenkin:2000jqa}, and the loop formation efficiency $C_{\rm eff} = 5.7$ in the radiation dominated universe \cite{Vanchurin:2005pa,Ringeval:2005kr,Olum:2006ix,Olmez:2010bi,Blanco-Pillado:2013qja,Blanco-Pillado:2017oxo,Cui:2018rwi}.
For QSS, the contribution from each normal mode of the loops formed before $t_{\rm fin}=t_M$ can also be given by Eq.~\eqref{eq:GWs2} with the terms involving $\frac{1}{2\pi R_c}$ removed. The contribution to the gravitational wave background from the $MS\bar{M}$ structures of QSS is computed using the burst method as described in Ref.~\cite{Lazarides:2022jgr}:
\begin{align}\label{eq:GWs-Omega-MSM}
\Omega_{GW}^{MS\bar{M}}(f) = \frac{4\pi^2}{3H_0^2}f^3\int_{z_*}^{z_M}dz \, h^2(f,z)\frac{dR}{dz} \ ,
\end{align}
where $H_0$ is the present Hubble parameter and the lower limit $z_*$ of the integral over the redshift $z$ which separates the infrequent bursts from the stochastic background is computed from
\begin{align}
\int_0^{z_*}dz \,\frac{dR}{dz} = f .
\end{align}
 The length of an $MS\bar{M}$ structure evolves as
\begin{align}
\tilde{l}(z) = 2t_M - \tilde{\Gamma}G\mu(t(z)-t_M), 
\end{align}
with  $\tilde{\Gamma}\sim 8\ln \gamma(z_M)$ and $\gamma \sim \frac{\mu}{m_M}\tilde{l}$ ($m_M$ is the monopole mass). The wave form for the gravitational wave burst is given by \cite{Martin:1996ea,Martin:1996cp,Leblond:2009fq}
\begin{align}
h(f,z)=\tilde{g}\frac{G\mu \tilde{l}(z)}{r(z)}f^{-1}  ,
\end{align}
where $\tilde{g}=2\sqrt{2}\sin^2\sqrt{2}/\pi$ and $r$ is the proper distance.
 The number density of the $MS\bar{M}$ structures is 
\begin{align}
\label{eq:density_mm_pair}
\tilde{n}(z)=(2t_M)^{-3}\left(\frac{1+z}{1+z_M}\right)^3 .
\end{align}
The burst rate can be written as
\begin{align}
\frac{dR}{dz} = \frac{4\pi r^2}{(1+z)^4H(z)}\frac{\tilde{n}(z)}{\tilde{l}(z)}\Delta(f,z) \ ,
\end{align}
where $\Delta(f,z)$ is the fraction of the gravitational wave bursts that are observed (see Ref.~\cite{Lazarides:2022jgr} for more details). 
Figs.~\ref{fig:GWs-QSS} and \ref{fig:GWs-DWS} respectively show the gravitational wave background from the QSS and the WBS structures in the case of superheavy strings with $G\mu = 10^{-6}$ and $t_M=3\times 10^2$ sec, and $G\mu = 10^{-6}$ with $v_{\rm dw}=10^2$ GeV, the vacuum expectation value associated with the walls.

The string network experiences some $e$-foldings during inflation and the loop formation starts around $t_F\sim 10^{-10}$ sec. This alleviates the bound from the advanced LIGO-VIRGO third run (LV3) \cite{LIGOScientific:2021nrg}. It is worth mentioning that non-standard cosmology such as matter domination (MD) in the pre-BBN era or both inflation and MD can also alleviate the LV3 bound around the decaHertz region \cite{Maji:2023fhv,Lazarides:2023rqf,Lazarides:2023bjd}. As an example, consider models with partial inflation of the strings 
starting right after their formation and followed by a matter  dominated period of field oscillations which ends by reheating. In 
this case, using Ref.~\cite{Lazarides:2021uxv} we find that the horizon reentry time is given by
\begin{align}
t_F\sim \xi_{\rm str}^2 \exp\left(2N_{\rm str}\right)t_r^{1/3}t_e^{-4/3} ,
\end{align}
where $\xi_{\rm str}$ is the correlation length at string formation, $N_{\rm str}$ the $e$-foldings experienced by the strings, $t_r$ the reheat time, and $t_e$ the time at which inflation terminates. The gravitational wave background predicted in other frequencies can be tested in various proposed experiments including HLVK \cite{KAGRA:2013rdx}, CE \cite{Regimbau:2016ike}, ET \cite{Mentasti:2020yyd}, DECIGO \cite{Sato_2017}, BBO \cite{Crowder:2005nr, Corbin:2005ny}, LISA \cite{Bartolo:2016ami, amaroseoane2017laser}, and SKA \cite{5136190, Janssen:2014dka}.

\begin{figure}[htbp!]
\begin{center}
\includegraphics[width=\linewidth]{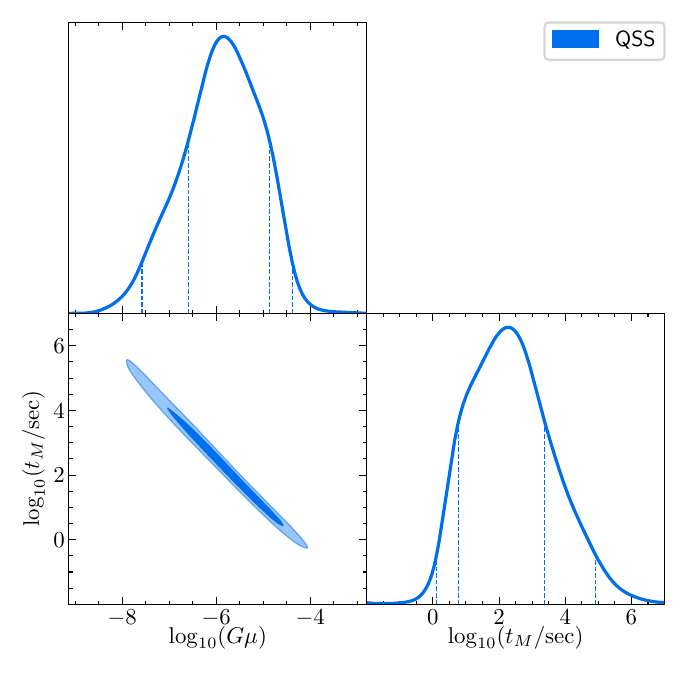}
\end{center}
\caption{Corner plot of the posterior distribution of $G\mu$ and $t_M$ for QSS with the $1\sigma$ (dark) and $2\sigma$ (light) credible regions. The diagonal plots are the marginalized 1D distributions with the vertical lines indicating the $68\%$ and $95\%$ confidence level intervals.}\label{fig:triang-QSS}
\end{figure}
\begin{figure}[htbp!]
\begin{center}
\includegraphics[width=\linewidth]{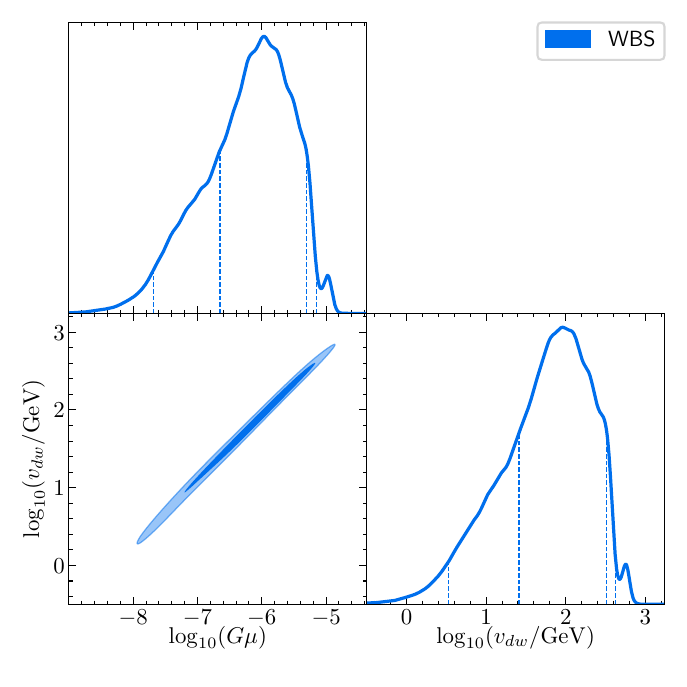}
\end{center}
\caption{Corner plot of the posterior distribution of $G\mu$ and $v_{\rm dw}$ for WBS with the $1\sigma$ (dark) and $2\sigma$ (light) credible regions. The diagonal plots are the marginalized 1D distributions with the vertical lines indicating the $68\%$ and $95\%$ confidence level intervals.}\label{fig:triang-WBS}
\end{figure}
We perform a Bayesian analysis of these two models using the wrapper PTArcade \cite{Mitridate:2023oar} and the NANOGrav 15 year data. We employ the \texttt{Ceffyl} package \cite{Lamb:2023jls} and take into account the quadrupolar Hellings-Downs (HD) correlation \cite{Hellings:1983fr} between the pulsars to obtain the posterior distributions of the model parameters for the stochastic gravitational wave background. To estimate the Bayes factor of the models with respect to the supermassive black hole binaries (SMBHB), we use the \texttt{Enterprise} code \cite{enterprise, enterpriseext} without the HD correction. The Bayes factors are estimated to be around 60 and 30 for QSS and WBS, respectively, which are comparable with the  Bayes factors for the metastable strings ($\sim 20$) and the superstrings ($\sim 50$) \cite{NANOGrav:2023hvm}. The posterior distributions of $G\mu$ and $t_M$ for QSS are shown in the triangular plot in Fig.~\ref{fig:triang-QSS}. The plots in Fig.~\ref{fig:triang-WBS} depict the posteriors of the model parameters $G\mu$ and $v_{\rm dw}$ of WBS.
\begin{table}
\begin{tabular}{lcccc}
\hline
\multirow{2}{*}{Model} & \multirow{2}{*}{Parameters} & \multicolumn{2}{c}{Credible Intervals} \\
\cline{3-4}
 & & 68\% & 95\% \\
\hline
\multirow{2}{*}{QSS} & $\log_{10}(G\mu)$ & $[-6.59, -4.87]$ & $[-7.58, -4.38]$ \\
 & $\log_{10}(t_M/\mathrm{sec})$ & $[0.78, 3.36]$ & $[0.12, 4.91]$ \\
\hline
\multirow{2}{*}{WBS} & $\log_{10}(G\mu)$ & $[-6.65, -5.31]$ & $[-7.68, -5.15]$ \\
 & $\log_{10}(v_{\rm dw}/\mathrm{GeV})$ & $[1.41, 2.52]$ & $[0.53, 2.63]$ \\
\hline
\end{tabular}
\caption{Bayesian credible intervals for the model parameters of QSS and WBS.}
\label{tab:params}
\end{table}

Table~\ref{tab:params} presents the $68\%$ and $95\%$ confidence level intervals of the model parameters for QSS and WBS.  We find that the stochastic gravitational wave background for QSS and WBS is compatible with the NANOGrav 15 year data for $G\mu\approx 10^{-6}$, and $t_M\approx 10^2$ sec and $v_{\rm dw}\approx 10^2$ GeV respectively.

For completeness, note that the $G\mu$ values of interest in this paper correspond to symmetry breaking scales $\sim 10^{15}$ - $10^{16}$ GeV.  In the example of WBS given in Eq.~\eqref{eq:f-su5}, the  proton decay ( $p \to e^+ \pi^0$) is mediated by superheavy leptoquark gauge bosons associated with the symmetry breaking responsible for string formation. The lifetime is estimated to be larger than the current lower bound provided by Super-Kamiokande \cite{Super-Kamiokande:2020wjk} for $G\mu\gtrsim 10^{-6}$. Proton decay can be observed within the next few years at Hyper-Kamiokande \cite{Dealtry:2019ldr} for $G\mu\lesssim 3\times 10^{-6}$.

\section{Summary}
\label{sec:summary}
Inspired largely by the apparent evidence of a low frequency gravitational wave background by the NANOGrav and other collaborations, our main aim here is to point how realistic extensions of the Standard Model of high energy physics can now be tested by this and hopefully near future discoveries. We have focused on two distinct composite topological structures, namely, quasi-stable cosmic strings and walls bounded by strings. The local strings in both cases are superheavy with $G\mu \approx 10^{-6}$ and the two composite structures are unstable and decay through the emission of gravitational waves. A certain amount of inflation of the strings is necessary in order that the gravitational emission is also compatible with the LIGO-VIRGO constraint on $G\mu$ in the $10-100$ Hz frequency range. The gravitational wave background from the quasi-stable strings with $\log_{10}(G\mu) = -5.84_{-0.75}^{0.97}$ and $\log_{10}(t_M/\mathrm{sec})=2.28_{-1.5}^{1.08}$, and walls bounded by strings with $\log_{10}(G\mu) = -5.97_{-0.68}^{0.66}$ and $\log_{10}(v_{\rm dw}/\mathrm{sec})=1.96_{-0.55}^{0.56}$ can explain the recent evidence in the NANOGrav 15 year data. The Bayes factors for QSS and WBS provide strong evidence for these scenarios in comparison with the simplest model for SMBHBs, and they are comparable with the other competing models of cosmic strings such as metastable strings and cosmic superstrings.

\begin{acknowledgments}
We thank Amit Tiwari for a careful reading of the manuscript. R.M. acknowledges discussions with Sunando Patra and Nilanjandev Bhaumik. The authors thank Nilanjandev Bhaumik for a run of the model files for QSS on his computer as a cross-check. This work is supported by the National Research Foundation of Korea grants by the Korea government: 2022R1A4A5030362 (R.M.) and the Hellenic Foundation for Research and Innovation (H.F.R.I.) under the “First Call for H.F.R.I. Research Projects to support Faculty Members and Researchers and the procurement of high-cost research equipment grant” (Project Number: 2251) (G.L. and Q.S.). R.M. thanks Rick Bernard and the Department of Physics \& Astronomy in the University of Delaware for providing access to Mathematica.
\end{acknowledgments}

\bibliographystyle{mystyle}
\bibliography{strings_aps_2}
\end{document}